# Using rectangular collocation with finite difference derivatives to solve electronic Schrödinger equation

Sergei Manzhos[a,1], Tucker Carrington Jr[b]

[a] Department of Mechanical Engineering, National University of Singapore, Block EA #07-08, 9 Engineering Drive 1, Singapore 117576 Singapore.
[b] Chemistry Department, Queen's University, Kingston, ONT. K7L 3N6, Canada

**Abstract**

We show that a rectangular collocation method, equivalent to evaluating all matrix elements with a quadrature-like scheme and using more points than basis functions, is an effective approach for solving the electronic Schrödinger equation (ESE). We test the ideas by computing several solutions of the ESE for the H atom and the $H_2^+$ cation and several solutions of a Kohn-Sham equation for CO and $H_2O$. In all cases, we achieve millihartree accuracy. Two key advantages of the collocation method we use are: 1) collocation points need not have a particular distribution or spacing and can be chosen to reduce the required number of points; 2) the better the basis, is the less sensitive are the results to the choice of the point set. The ideas of this paper make it possible to use *any* basis functions and thus open to the door to using basis functions that are not Gaussians or plane waves. We use basis functions that are similar to Slater type orbitals. They are rarely used with the variational method, but present no problems when used with collocation.

## 1 Introduction

Quantum chemistry methods are important in chemical physics because by solving the electronic Schrödinger equation (ESE), one can determine the structure of molecules, reaction pathways, reaction barrier heights etc.[1] The ESE is usually solved by using a variational method and solving a matrix eigenvalue problem.[2] In this paper, we demonstrate that it is possible to solve the ESE with a collocation-like method used previously to solve the vibrational Schrödinger equation.[3-13] The success of the method is based on the fact that if all (also those of the kinetic energy operator,(KEO) and the overlap matrix) matrix elements are evaluated by quadrature, then accurate results are obtained for all states whose wavefunctions are in the space spanned by the basis.[14] This is true even if the quadrature error is not small. It is crucial that there be more points than basis functions. Other authors have used collocation-type methods. The method of Friesner is the most successful.[15-17] It differs from our method in two key ways. 1) To compensate for incompleteness of the basis, in the approach of Friesner, one uses a (weighted) least squares method with de-aliasing functions to transform a vector of coefficients from a grid to a basis. On the other hand, we deal with incompleteness of the basis by following Boys[14] and Nakatsuji,[18,19] using equations for the matrix elements that are formally quadrature equations. 2) The method of Friesner calculates matrix elements of the standard one-electron Hamiltonian and the overlap matrix analytically. Instead, *all* of our matrix elements are computed by summing over grid points (equivalent to quadrature). A significant advantage of any collocation-like method is the ability to use *any*, i.e. not Gaussian and not plane-wave, basis functions. This may make it possible to compute converged levels with fewer basis functions than would be required with a Gaussian basis. This advantage is a consequence of not relying on analytic integrals. Early in the history of quantum chemistry, Slater type basis functions were abandoned because of the difficulty of the associated integrals.[20] Collocation and ideas of Boys[14] make it possible to use the best possible basis functions without regard for the difficulty of integrals.

---

[1] Author to whom correspondence should be addressed. E-mail: mpemanzh@nus.edu.sg ; Tel: +65-6516-4605.

Most electronic structure calculations use one of two types of bases: localized, atom-centered functions, or plane waves.[21,22] The plane wave basis sets provide the convenience of a generic functional form with a single parameter controlling the accuracy – the cutoff frequency (often called the cutoff energy).[23,24] They are naturally suited for periodic systems and do not suffer from basis set superposition error (BSSE).[23] The main disadvantages of a plane wave basis are: 1) its large size and the associated memory cost; 2) its intrinsic periodicity which means that for a non-periodic system, calculations must be performed in large supercells; and 3) the slow convergence of the basis expansion when the wavefunction (i.e. cusp) or density is rapidly changing, necessitating the use of pseudopotentials.[25] Although the pseudopotentials significantly reduce the CPU cost, they are a source of error. With plane wave basis sets, the required matrix is large, and matrix elements of all terms except the KEO and the overlap can be computed by quadrature or in Fourier space (by expanding, e.g. the pseudopotential). The large plane wave basis implies a large equivalent and equidistant real space quadrature grid on which the potential is evaluated. Because KEO and overlap matrix elements are computed exactly, quadrature for potential terms must be accurate and the grid usually covers all space where wavefunctions have appreciable amplitude. A large grid size is also required in finite-difference based approaches.[26]

Localized basis sets are often based on Gaussian or Gaussian like (e.g. determined numerically) functions centered on ions.[27] With localized basis functions, the required matrix size is much smaller than with plane waves, and with Gaussian functions, all matrix elements can be computed analytically. Even when the matrix size is large (owing to a large number of electrons), localized basis functions are advantageous; there are linear-scaling formulations of density functional theory (DFT)[28] using localized basis functions which are strictly zero outside a defined region of space. The relatively small size of the matrix, using Gaussian or Gaussian-like functions, is due the fact that the atom-centered basis functions mimic atomic states; good results can be obtained with a small basis set because linear combinations of a few atomic states provide *qualitatively* reasonable approximations to electronic states in many compounds.[29] This reduces memory and CPU costs and allows for direct diagonalization. To achieve converged energy levels, however, the basis needs to be expanded considerably, which increases the CPU cost and may create problems due to the conditioning of the overlap matrix. For example, the 6-31g(d,p) basis, which is popular in DFT modeling of organic molecules, results in orbital energies which are about 0.3 eV overestimated compared to those computed with a more complete 6-31+g(d,p) basis (i.e. adding the so-called diffuse functions). Atom-centered bases also suffer from BSSE. An effective way to improve accuracy without increasing the basis size is to tune the shape of the functions. Tuning is manpower-costly and more involved than the relatively simple choice of increasing the plane wave cutoff, but allows obtaining an accuracy similar to that achieved with a large plane wave basis. For example, we were able to achieve interaction energies of similar accuracy to those obtained with a plane wave basis by using numerical DZP (double-ζ polarized) bases tuned using simple rules, even though the default basis sets of small enough size to make calculations routine are often not accurate.[30-32] Radial parts of atom-centered bases sets are carefully chosen to fit atomic states;[28,33,34] this allows performing all-electron calculations. The quality of the basis depends critically on the quality of the radial part. Non-equidistant, albeit symmetric, grids around ions are often used.[35] Although very useful for many molecules, the advantage of having atom-centered basis functions is less important when they are very mixed; metallic bonding is an example. In this paper, we use atom-centered basis functions, but not Gaussians.

Both atom-centered and plane-wave bases are usually used with a variational method. This is true regardless of whether one solves the ESE, the Kohn-Sham (KS) equation or the Hartree-Fock equation. In a variational calculation, one applies operator $\hat{O} - E$ to a basis expansion and demands that the resulting function be orthogonal to all functions in the basis, where $\hat{O}$ is the Hamiltonian or the Fock operator, or the KS operator, and *E* is an energy to be determined. The resulting matrix eigenvalue problem is solved with methods of numerical linear algebra. The eigenvalues approach the exact levels from above as the basis size is increased.[36] A disadvantage of this variational approach is that when some but not all matrix elements are computed by quadrature, the accuracy of the levels one computes depends on the accuracy of the quadrature. Even if quadrature points and weights, with which one can accurately compute integrals of basis functions, are known, matrix elements may not be accurate.

Collocation is an alternative formulation.[37-39] We solve, for example, the Schrödinger equation (we use atomic units)

$$-\frac{1}{2}\Delta\psi(\pmb{x}) + V(\pmb{x})\psi(\pmb{x}) = E\psi(\pmb{x}),$$

(1)

where $\psi(\pmb{x})$ is the wavefunction and $V(\pmb{x})$ the Coulombic potential due to the ions and electrons.[37-39] The same ideas work when the potential is the effective KS potential and $\psi(\pmb{x})$ is a KS single-electron orbital and when the operator on the left is the Fock operator. Collocation requires that the SE be satisfied at a set of points $\{\pmb{x}_i\}, i = 1, \ldots, M$. Using a basis expansion $\psi(\pmb{x}) = \sum_{k=1}^{N} c_k \phi_k(\pmb{x})$:

$$-\frac{1}{2}\sum_{k=1}^{N} c_k \Delta\phi_k(\pmb{x}_i) + V(\pmb{x})\sum_{k=1}^{N} c_k \phi_k(\pmb{x}_i) = E \sum_{k=1}^{N} c_k \phi_k(\pmb{x}_i)$$

(2)

Or in the matrix form

$$\pmb{Dc} + \pmb{VFc} = E\pmb{Fc},$$

(3)

where $D_{ik} = -\frac{1}{2}\Delta\phi_k(\pmb{x}_i)$, $F_{ik} = \phi_k(\pmb{x}_i)$, $V_{ik} = \delta_{ik}V(\pmb{x}_i)$, and $\pmb{c}$ is vector. The elements of the matrix $\pmb{D}$ can be computed analytically[4-6,8-12,40,41] or numerically.[3,7,13] In the latter case, collocation is easily used with *any* basis functions. Moreover, there is no requirement of smoothness of the basis functions other than at collocation points; there is also no requirement that $\phi_k(\pmb{x})V(\pmb{x})\phi_{k\prime}(\pmb{x})$ be integrable. Equation 3 is a rectangular matrix equation because in general $M > N$. It can be solved in the least squares sense as a rectangular matrix pencil.[4,5,11,42] Alternatively, it can be converted to a square generalized eigenvalue problem by left multiplying by $\pmb{F}^T$ [3,5-7,13,14,18,19]

$$\boldsymbol{F}^T(\boldsymbol{D} + \boldsymbol{VF})\boldsymbol{c} = E\boldsymbol{F}^T\boldsymbol{F}\boldsymbol{c}. \qquad (4)$$

The same equation is solved when one does a variational calculation and computes *all* the matrix elements using quadrature with unit weights. However, eigenvalues of Eq. 4 may be accurate even if that quadrature is poor. Indeed, Eq. 4 works even if Hamiltonian matrix elements are infinite. We have shown when solving the vibrational SE (which has the same form as Eqs. 1-3) that there is considerable freedom in the choice of $\{\boldsymbol{x}_i\}$, e.g. choosing to give more weight to certain regions of space (such as the bottom of the potential well). In this way, accurate spectra can be obtained with modestly-sized sets of collocation points $\{\boldsymbol{x}_i\}$. In a variational calculation, because matrix elements of the KEO and the overlap matrix are exact, it is necessary to have quadrature points (for the potential) in the region in which wavefunctions have amplitude. On the other hand, in a collocation calculation, because *all* matrix elements are evaluated by summing over grid points, it is sufficient to have points only in part of that region. Using more points than basis functions much improves the accuracy of the energy levels we compute, but forces us to deal with a rectangular problem. Putting $\boldsymbol{F}^T$ on the left gives us a least squares solution to the rectangular problem.[5,7] Collocation has been used to solve high dimensional vibrational problems by using Smolyak grids,[43,44] which have structure that greatly facilitates the evaluation of matrix-vector products. A disadvantage of collocation is the fact that the matrix $\boldsymbol{F}^T(\boldsymbol{D} + \boldsymbol{VF})$ is not symmetric which for some eigensolvers increases the cost of computing the eigenvalues. Eigenvalues that are not converged may be complex.

Although collocation-like methods are becoming more popular for the vibrational SE,[3-13,43,44,45,46] they are not commonly used for the solution of the electronic problem. Nakatsuji et al. used the so-called local Schrödinger equation (LSE), which is a type of a collocation equation, with the free-complement method, to obviate the need to compute Hamiltonian matrix elements.[18] They demonstrated the advantage of non-uniform point placement,[19] weighting more regions with the largest wavefunction amplitude, for computing the *ground* state of few-electron systems. When solving the vibrational SE, we emphasized the value of sampling according to the value of the potential and this was found to be effective for computing *multiple* levels. Friesner and co-workers have extensively used a pseudo-spectral method to solve the electronic Schrödinger equation. It is related to collocation. An important advantage of the approach of Boys that we use is that it can easily deal with more points than basis functions and there is no need to compensate for aliasing errors. Anderson and Ayers have used a Boys-type method to solve the ESE; they use a Smolyak grid.[47,48] McCormack et al. have used Boys idea to test quadrature grids.[49] They use standard grids and attempt to reduce the number of points.

*The purpose of this paper is to study the applicability of the rectangular collocation approach with a numerical KEO to the solution of the electronic Schrödinger equation using atom-centered basis sets.* We consider in this study several systems which allow us to test key ideas. All calculations are in three dimensions, both for simplicity and because most quantum chemistry calculations require solving either the Hartree-Fock or the KS equation which are in 3D. We solve the ESE for the hydrogen atom and the $H_2^+$ ion. We solve the KS equation for CO and $H_2O$. In all cases, we compute several of the lowest energy levels. We test our solutions for H and $H_2^+$ versus known exact levels and for CO and $H_2O$, versus reference DFT calculations which use the variational approach.

## 2    Methods

The Schrödinger Eq. 1 was solved using the Ansatz of Eqs. 2-4 with $\boldsymbol{x} = \{x, y, z\}$. Elements of the matrix $\boldsymbol{D}$ are computed with a five-point finite-difference (FD) stencil, with a FD step of $1 \times 10^{-6}$ *a.u.* Results computed with stencils with fewer points were less accurate and using stencils with more than five terms did not noticeably improve the accuracy. The calculations were performed in Octave;[50] the generalized eigenvalue problem was solved using the *eig* function. The quality of the solution was evaluated vs. the reference levels and by monitoring the residual

$$Res \equiv \frac{\langle|\psi|||H\psi - E\psi|\rangle}{\langle\psi|\psi\rangle} \approx \frac{|\boldsymbol{c}^T\boldsymbol{F}^T||(\boldsymbol{D} + (\boldsymbol{V} - E)\boldsymbol{F})\boldsymbol{c}|}{\boldsymbol{c}^T\boldsymbol{F}^T\boldsymbol{F}\boldsymbol{c}}$$

(5)

for relevant levels. Note that in Eq. 5 "| |" denotes absolute values of vector elements or function values and not the norm. We use the residual of Eq. 5 because it is a rectangular residual vector, $(\boldsymbol{D} + (\boldsymbol{V} - E)\boldsymbol{F})\boldsymbol{c}$, weighted with the absolute value of the wavefunction.

The basis functions were of the form $\phi_{k\zeta_{lm}}(\boldsymbol{x}) = \phi_\zeta(R_k)Y_{lm}(\boldsymbol{x} - \boldsymbol{x}_k)$, where $Y_{lm}$ is a spherical harmonic, $R_k = \|\boldsymbol{x} - \boldsymbol{x}_k\|$, and $\boldsymbol{x}_k$ is the 'center' of the $k^{th}$ set of basis function. The functions were centered on the ions. Naturally, multiple basis functions are positioned at each $\boldsymbol{x}_k$, and their number is controlled by a parameter $l_{max}$ such that $l = 0, 1, \ldots, l_{max}$, and $m = -l, -l+1, \ldots, l$, as is the case for the exact H atom solutions. We did calculations with several forms of the radial factor: a simple exponential function $\phi_\zeta(R) = e^{-\epsilon_\zeta R}$; the Gaussian function $\phi_\zeta(R) = e^{-\epsilon_\zeta R^2}$; the Matern3/2 function $\phi_n(R) = e^{-\epsilon_\zeta R}(1 + \epsilon_\zeta R)$; the Matern5/2 function $\phi_\zeta(R) = e^{-\epsilon_\zeta R}(1 + \epsilon_\zeta R + \frac{1}{3}(\epsilon_\zeta R)^2)$; and the inverse multiquadratic function $\phi_\zeta(R) = (1 + \epsilon_\zeta R^2)^{-b/2}$ with different $b$.[51] The simple exponential $\phi_\zeta(R)$ was the best and we give only results for it. The exponential form has the advantage that it has the correct form close to the nuclear cusp;[52] however, other functions might be advantageous when the potential is smoother e.g. when using pseudopotentials. Slater type basis functions (close to the simple exponential) are not popular in quantum chemistry because Gaussian basis functions facilitate the evaluation of integrals. Collocation-type methods do not require integrals. The parameter $\epsilon_\zeta$ determines the basis function width. At each center, we use several $\epsilon_\zeta$ which we denote, $\zeta = 1,\ldots, \zeta_{max}$. This is what is commonly known as a multi-zeta basis.[2] The values of $\epsilon_\zeta$ were chosen approximately to minimize *Res*.

Collocation points $x_i$ were selected from a uniform grid filling a $L_x \times L_y \times L_z$ cuboid centered on the ion(s) and having $N_x$ points along $x$ etc. The 1D densities, $N_a/L_a$, $a = x, y, z$, are equal and varied to achieve accurate energies. Except for $H_2^+$, where the Schrödinger equation was solved for a range of interatomic distances, the lengths of the sides were equal and denoted by $L$. In the case of $H_2^+$, one of the sides was extended by the value of the interatomic distance $r_{HH}$. Points are selected from the cube or cuboid with a bias towards low values of the potential, as was done previously for vibrational problems. A grid point $x_i$ is accepted into the collocation point set if[3,7,13,53]

$$\frac{V_{max} - V(x_i) + \delta}{V_{max} - V_{min} + \kappa} > rand$$

(6)

where *rand* is a random number in the range [0, 1]. We will show below that the shape of electronic potentials makes point selection more difficult than in the vibrational case. The parameters $\delta, \kappa$ were therefore used to adjust the point distribution to ensure the accuracy of all computed levels. Because of the random component of point selection, there is slight variability of results from run to run. However, all runs give energies with similar errors.

For the H atom and the $H_2^+$ ion, the potential $V(x)$ in Eq. 1 is,

$$V_H(x) = -\frac{1}{\|x - x_H\|}$$

$$V_{H_2^+}(x) = -\frac{1}{\|x - x_{H_1}\|} - \frac{1}{\|x - x_{H_2}\|}.$$

(7)

For the DFT calculations,[22] Eq. 1 becomes

$$-\frac{1}{2}\Delta\psi(x) + V_{eff}(x)\psi(x) = E\psi(x)$$

(8)

where

$$V_{eff}(x) = V_{nuc}(x) + \int \frac{\rho(x')}{|x-x'|}dx' + V_{XC}(x),$$

(9)

where $\rho(x)$ is the electron density and $V_{XC}(x)$ is the exchange-correlation potential. The sum of the first two terms in Eq. 9 is the total electrostatic potential (ESP). Both the ESP and the density are output by Gaussian 09.[33] The 6-311++g(2d,2p) basis was used. We used the Xα functional[54] for simplicity of constructing the exchange-correlation potential (which in the case of Xα is exchange-only):

$$V_x(x) = 0.7\frac{3}{2}\left(\frac{3}{\pi}\right)^{\frac{1}{3}}\rho(x)^{\frac{1}{3}}$$

(10)

Obviously, any exchange-correlation functional could be used; we chose Xα for these first tests of our collocation approach to avoid possible inaccuracies when extracting $V_{eff}$ from a DFT code.

## 3  Results

*3.1  Schrödinger equation for the hydrogen atom*

**Table 1** lists the exact H atom energy levels ($E_n = -0.5n^{-2}$, where *n* is the principal quantum number), and levels computed with collocation, for different box sizes *L* and numbers of points in each dimension, $N_x = N_y = N_z$. The values of $\epsilon_\zeta$ were manually adjusted to approximately minimize *Res* (Eq. 5) for the first 14 levels up to 3*d*. We used $\epsilon_1 = 0.3$, $\epsilon_2 = 0.39$, $\epsilon_3 = 0.54$, $\epsilon_4 = 0.63$, $\epsilon_5 = 0.96$. We note that the maximum value of the Coulomb potential in the cube, $V_m \approx -\frac{2}{\sqrt{3}L}$, is, for some *L* values, below some of the energy levels we compute, highlighting the fact that it is possible to use collocation points in a region that does not extend to the turning points. We have no points in the exponential tails of the wavefunctions. Moreover, after using Eq. 6 to determine points, we discard all points below -10 *a.u.* and we therefore have no points close to the Coulombic singularity.

**Table 1**. H atom exact energy levels and levels computed with collocation, in *a.u.*, for different $L$ and $N_x = N_y = N_z$ values and different bases and collocation point sets.

|  | Exact | collocation | | | | |
| --- | --- | --- | --- | --- | --- | --- |
| state |  | Energy | | | | |
| 1s | -0.5000 | **-0.4999** | -0.4862 | -0.500 | -0.4668 | -0.4998 |
| 2s, 2p | -0.1250 | **-0.1251** | -0.1250 | -0.1241 | -0.1250 | -0.1252 |
|  |  | **-0.1251** | -0.1248 | -0.1241 | -0.1249 | -0.1251 |
|  |  | **-0.1250** | -0.1247 | -0.1239 | -0.1248 | -0.1251 |
|  |  | **-0.1249** | -0.1247 | -0.1225 | -0.1248 | -0.1248 |
| 3s, 3p, 3d | -0.0556 | **-0.0557** | -0.0563 | -0.0566 | -0.0560 | -0.0550 |
|  |  | **-0.0554** | -0.0563 | -0.0566 | -0.0559 | -0.0547 |
|  |  | **-0.0553** | -0.0561 | 0.0564 | -0.0558 | -0.0542 |
|  |  | **-0.0551** | -0.0561 | -0.0563 | -0.0558 | -0.0536 |
|  |  | **-0.0547** | -0.0557 | -0.0562 | -0.0557 | 0.0236 |
|  |  | **-0.0544** | -0.0555 | -0.0561 | -0.0557 | 0.0272 |
|  |  | **-0.0544** | -0.0554 | -0.0557 | -0.0557 | -0.0353 |
|  |  | **-0.0539** | -0.0554 | -0.0557 | -0.0556 | 0.1079 |
|  |  | **-0.0536** | -0.0548 | -0.0555 | -0.0554 | 0.5219 |
| | $L$, a.u. | **30** | 30 | 15 | 30 | 30 |
| | Grid / Eq. 8 | **Eq. 8** | Eq. 8 | Eq. 8 | Grid | Eq. 8 |
| | $N_x$ | **50** | 50 | 50 | 50 | 50 |
| | $l_{max}$ | **d** | d | d | d | p |
| | $\zeta_{max}$ | **5** | 4 | 4 | 4 | 4 |
| | $N_{fns}$ | **45** | 36 | 36 | 36 | 20 |
| | $M$ | **3937** | 3872 | 3499 | 125000 | 3823 |

There is no need to worry about doing integrals of singular functions. It is only necessary that the basis be nearly complete in the vicinity of the collocation points.[14,7] In our calculation, levels that should be exactly degenerate are merely close. We report the actual values. Differences between nearly degenerate levels are a realistic measure of the error of the calculations. Symmetry could be easily imposed. **Table 1** shows that millihartree accuracy is easily achieved with a small number of collocation points and basis functions. For the 14 levels up to 3$d$, naturally, $d$ functions are required in the basis, and millihartree accuracy requires $\zeta_{max}$ = 5. It is possible to obtain energies of roughly comparable accuracy, without selecting points from the grid with Eq. 6, but to do so one needs almost two orders of magnitude more points (cf. energies in the second rightmost column of the table). The results therefore show that using more points in low-energy regions works well and allows obtaining good accuracy with many fewer points than would be required with a regular grid.

### 3.2 Schrödinger equation for the $H_2^+$ ion

We solved the Schrödinger equation for $H_2^+$ at several interatomic distances. The results for the 14 lowest states with the $l_{max} = 1$ basis, which we refer to as the *sp* basis, and the $l_{max} = 2$ basis, which we refer to as the *spd* basis, are shown in **Figure 1** and **Figure 2**. They are shown in comparison with the reference data of Ref. 55, which were tabulated from the analytic solution. The data are plotted by subtracting out the ion-ion interaction energy ($V_{ion} = 1/r_{HH}$) to obtain the potential energy curves (as in Ref. 55). The collocation point set was drawn from a uniform grid (of size (30+$r_{HH}$) ×30×30 Bohr with 50 points in each direction) using Eq. 6 with $\kappa = 0, \delta = 0.01\ a.u.$. The size of the set was about 8,000 (which differs slightly from run to run due to the random component of Eq. 6). Larger point sets lead to only slight improvements, and the accuracy degrades noticeably when $M$ < 5,000. The *sp* basis has 40 basis functions and the *spd* basis, 90 functions. Both bases used $\zeta_{max} = 4$ with the same values of the corresponding $\epsilon_\zeta$ as for H. When $\zeta_{max}$ < 4, we were not able to achieve millihartree accuracy. However, with a better optimization of the $\epsilon_\zeta$ values (we adjust manually) and/or an optimization of the form of the radial factors, it might be possible to achieve millihartree accuracy with a triple-$\zeta$ basis.

The *sp* basis appears to be sufficient to reproduce the lowest lying states, but it qualitatively fails for some higher-lying states, as expected. For the lowest states, it is excellent. For example, at the equilibrium $r_{HH}$, the ground and the first excited state are computed at -0.6029 and -0.1703 *a.u.* vs. the exact values of -0.6026 and -0.1675 *a.u.* With the addition of *d* type basis functions, millihartree accuracy is obtained for all the computed states, compare **Figure 2** to **Figure 1**. For example, at the equilibrium $r_{HH}$ the ground and the first excited states are computed at -0.6022 and -0.1697 *a.u.*. It is important that there are no fluctuations as a function of $r_{HH}$.

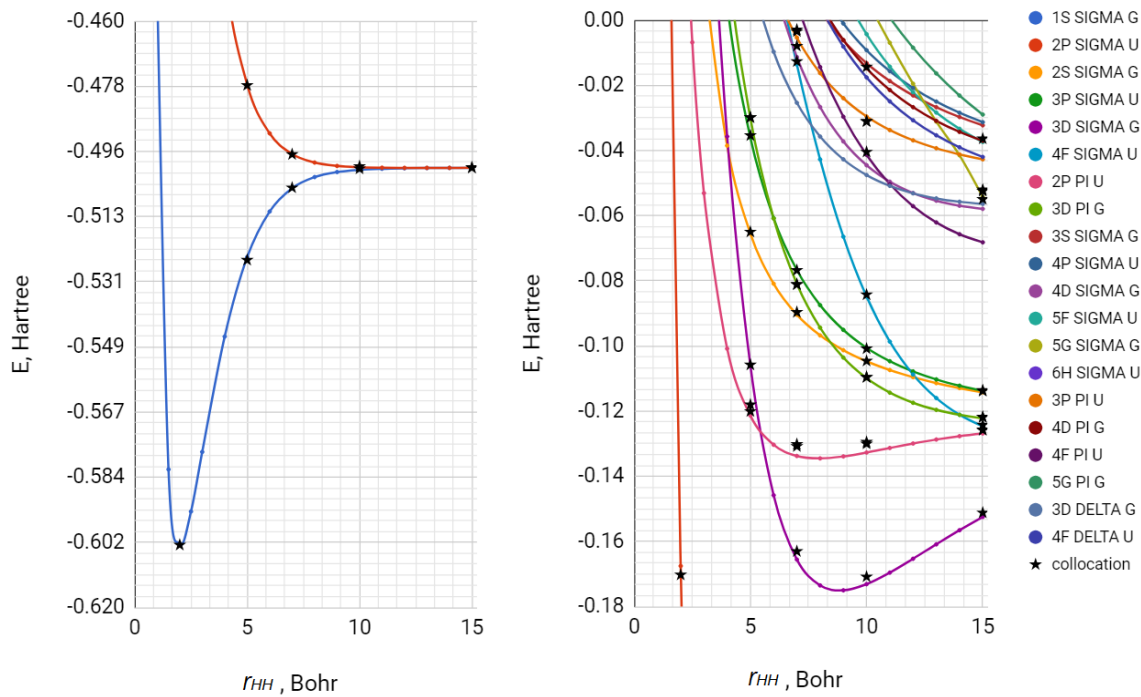

**Figure 1**. Potential energy curves for electronic states of $H_2^+$ from reference data of Ref. 55, and energies of the lowest 14 states obtained with collocation, using a basis with $l_{max} = 1$ (the *sp* basis). The lowest two electronic states are on the left and higher states are on the right. State labels follow Tables I-X of Ref. 55.

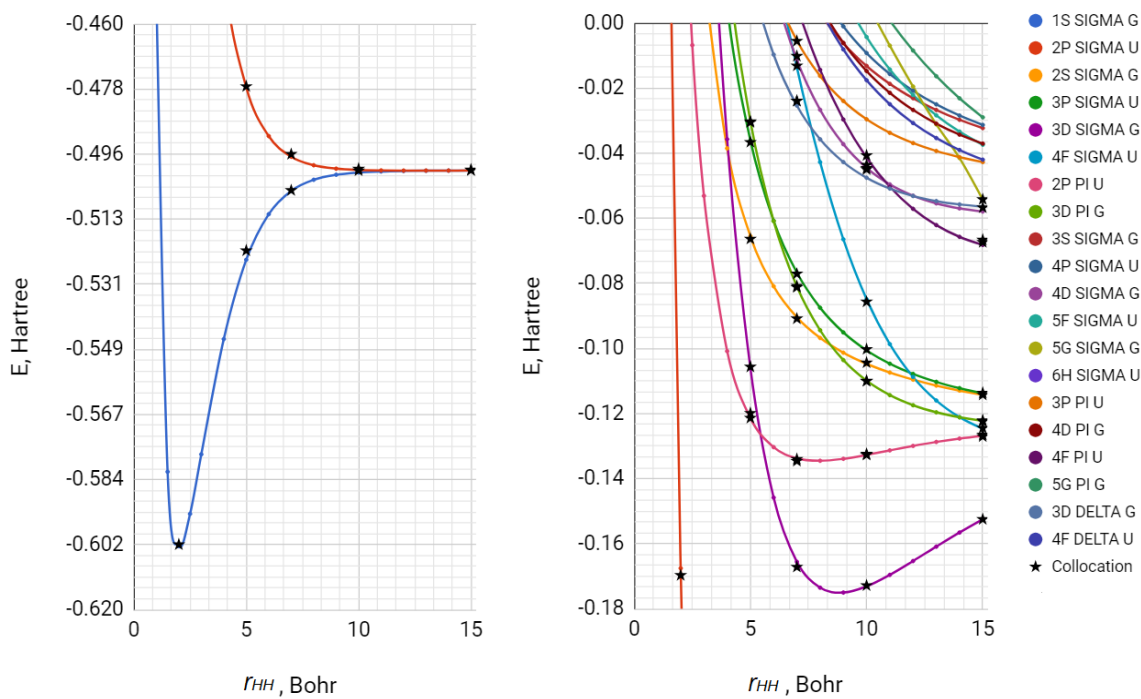

**Figure 2**. Potential energy curves for electronic states of $H_2^+$ from reference data of Ref. 55, and energies of the lowest 14 states obtained with collocation, using a basis with $l_{max} = 2$ (the *spd* basis). The lowest two electronic states are on the left and higher states are on the right. State labels follow Tables I-X of Ref. 55.

*3.3 The Kohn-Sham equation for CO and H$_2$O molecules*

For both molecules, we solved the KS equation at a DFT equilibrium geometry. The C-O bond length was set to 1.128 Å; the H-O bond length to 0.965 Å, and the HOH angle to 104 degrees. Electrostatic potentials were output from single-point Gaussian 09 calculations as cube files with a resolution of 0.0945 Bohr and 200 points in each direction. Cross-sections of the ESP are shown in **Figure 3** for CO and in **Figure 4** for H$_2$O. The plotted functions have sharp yet finite (due to the inclusion of the Hartree potential as well as to finite resolution) peaks at the atoms. The shape of the ESP is more complex than for the hydrogen atom and H$_2^+$ because different nuclei have different well depths. This makes collocation point selection with Eq. 6 more difficult than for H, H$_2^+$, and single-well vibrational problems, where simple overweighting of low-energy regions works well. Furthermore, for DFT molecular calculations, one wishes all of the levels up to and including those of the so-called frontier orbitals, rather than merely the lowest-energy levels. Some of these levels are deep in the two wells (of different depths) and some are near the top of the double well (close to the *x* axis in the figure). This poses difficulties for the choice of the basis. The lowest eigenvalues, in the core region, are widely spaced and correspond to core electronic states localized on individual ions, whereas the highest eigenvalues, in the valence region, are dense and correspond to states delocalized over the molecule. If an iterative eigensolver were used, the large spectral range and the disparate eigenvalue gaps would slow convergence. In this paper, we do not attempt to determine systematically the best point selection scheme. Instead, we use Eq. 6 with non-zero $\kappa$ and $\delta$ parameters, which results in a flat probability distribution in the region between $V_{min}$ and $V' = V_{min}+\kappa$ and in a simple potential value weighting for $V > V'$. We used $V' = -35$ *a. u.*, $\delta = 0.2$ *a. u.* for both molecules. These choices resulted in about 51,000 points drawn from the grid (of size 200$^3$). We note that the size of the original grid from which the collocation points are drawn is similar to typical sizes of grids used in plane-wave or finite difference based DFT calculations.[23,24] We thus use a small fraction of that number. There is of course no need to start with a regular grid, this was done here because Gaussian 09 outputs the ESP on a regular grid. Collocation points can be distributed in any desired way. Another consequence of the need to compute core and valence levels is a larger $\zeta_{max}$.

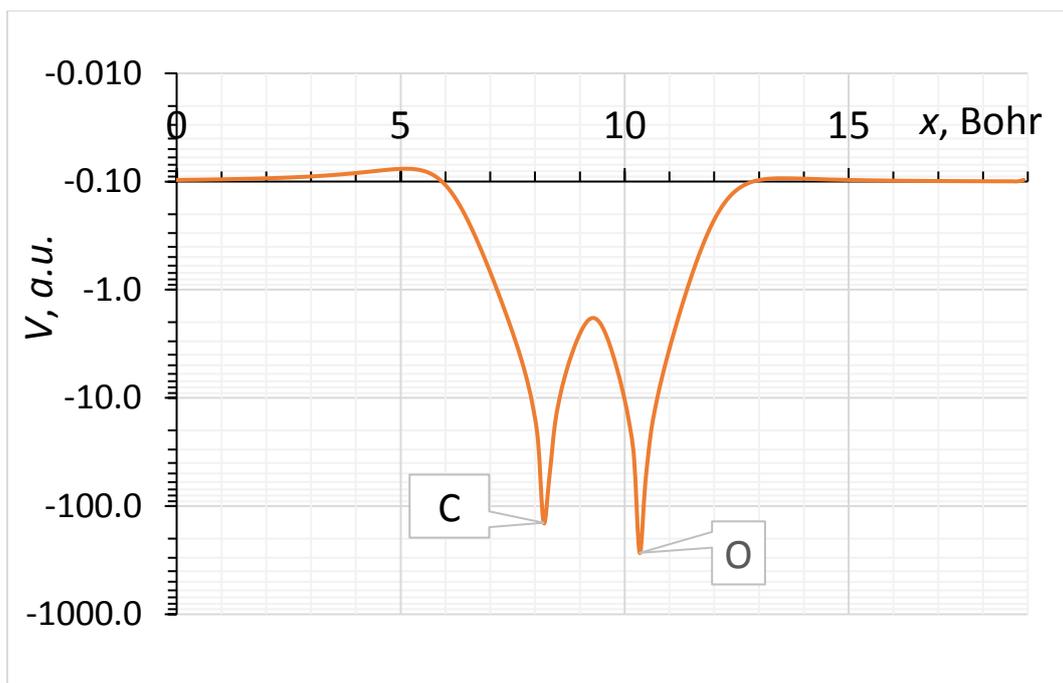

**Figure 3**. The profile of the electrostatic potential of CO along the molecular bond. Positions of the atoms are indicated by the callouts. Note the logarithmic scale and the shift of the origin by 0.1 Bohr.

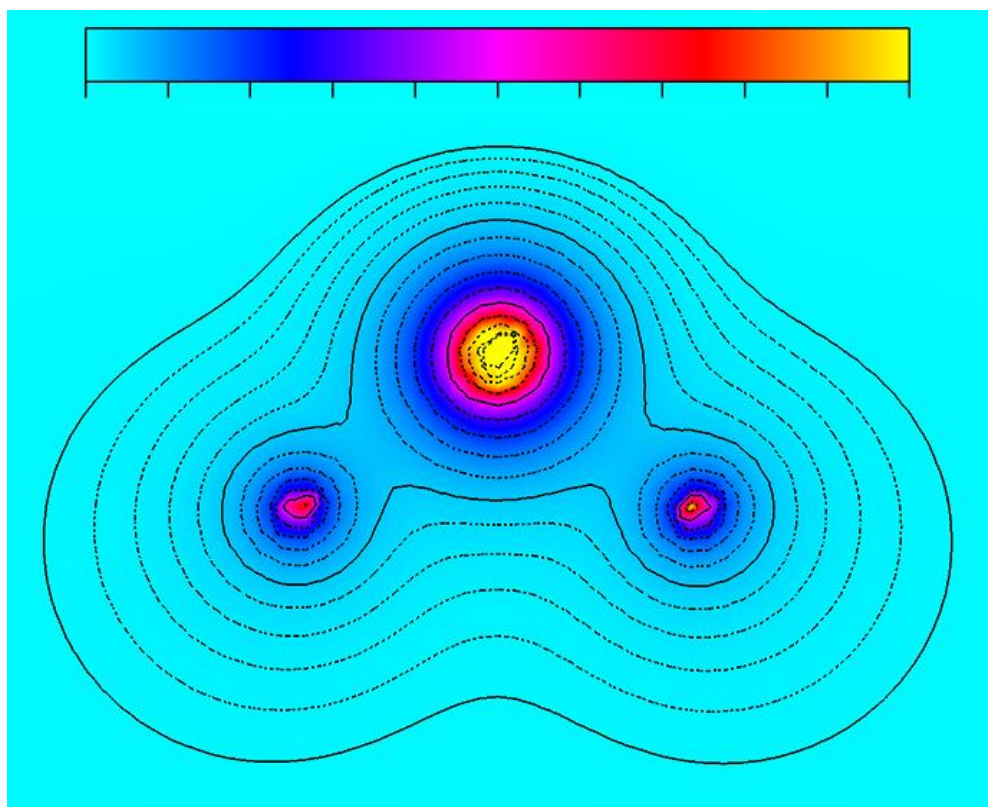

**Figure 4**. A 2D slice of the electrostatic potential of $H_2O$ along the molecular plane. Logarithmic scale with saturation at 20 *a.u*. Cyan to yellow color progression corresponds to going from the lowest to the highest values.

We computed the lowest energy levels starting from core 1s levels and through the molecular valence states. To get millihartree accuracy for CO, seven $\zeta$-components were required as well as $l_{max} = 2$ (an *spd* basis). The $\epsilon_\zeta$ values were 0.15, 0.195, 0.270, 0.315, 0.480, 0.810, and 1.350, multiplied by Z, where Z is nuclear charge. For $H_2O$, six $\zeta$-components were required and $l_{max} = 2$ (an *spd* basis). The $\epsilon_\zeta$ values for O were 0.30, 0.39, 0.54, 0.63, 0.96, and 1.62, all multiplied by Z, and 0.60, 0.78, 1.08, 1.26, 1.92, 3.24, for H. These parameters were manually selected to approximately minimize *Res*. The basis size was 126 for CO and 162 for $H_2O$. Adding *f* functions to the basis (basis size 224 for CO and 288 for $H_2O$) somewhat lowers the error. The Kohn-Sham eigenvalues obtained with the *spd* basis are listed in **Table 2** and compared to the reference values output by Gaussian 09. Millihartree accuracy is obtained. We note that deviations from the reference values are smaller than differences in the Kohn-Sham eigenvalues between different DFT programs using different or the same types of basis functions.[28,33] The deviation between the values computed by Gaussian 09 and with collocation is large for positive eigenvalues, but they have no physical meaning.

**Table 2**. Kohn-Sham eigenvalues for CO and $H_2O$ obtained with Gaussian 09 and with the collocation method (using an *spd* basis). All values are in *a.u.* The highest occupied and lowest unoccupied levels (HOMO and LUMO energies) are in bold.

| CO | | $H_2O$ | |
| --- | --- | --- | --- |
| Gaussian 09 | collocation | Gaussian 09 | collocation |
| -18.7401 | -18.7384 | -18.6306 | -18.6329 |
| -9.9072 | -9.9037 | -0.8931 | -0.8952 |
| -1.0480 | -1.0410 | -0.4521 | -0.4524 |
| -0.4912 | -0.4922 | -0.3140 | -0.3210 |
| -0.4133 | -0.4162 | **-0.2378** | **-0.2305** |
| -0.4133 | -0.4142 | **-0.0138** | **-0.0170** |
| **-0.3043** | **-0.3045** | 0.0391 | 0.0194 |
| **-0.0518** | **-0.0529** | 0.1407 | 0.0239 |
| -0.0518 | -0.0519 | | |
| 0.0384 | 0.0619 | | |

4  **Discussion and conclusions**

In this paper, we have shown that accurate solutions of the electronic Schrödinger equation and the Kohn-Sham equation can be obtained using a rectangular collocation method. Owing to the fact that all matrix elements are computed by summing over collocation points, if the basis set is sufficiently complete, it is not necessary to choose points and weights with which accurate integrals can be obtained.[14] The equations we solve have the same form as those of Boys but there is no need that the sums over points be considered quadrature approximations to integrals. One key advantage of the approach is that there is no need to choose basis functions that facilitate the calculation of integrals. It is only necessary to evaluate basis functions and their derivatives (computed with finite difference) at points. Gaussian and plane-wave bases are used in almost all electronic structure calculations because they facilitate the evaluation of integrals. When collocation is used, this advantage becomes moot. We have tested the ideas on the H atom, the $H_2^+$ molecular ion, as well as on the Kohn-Sham equation for CO and $H_2O$. In all these cases, we achieved millihartree or so-called chemical accuracy for several of the lowest-energy states with simple and small basis sets with about 100 functions and $10^3$-$10^4$ collocation points. We are not claiming that collocation is always the best way to solve electronic problems, but our exploratory calculations do demonstrate that this new tool should be further studied.

The rectangular collocation method has key advantages. 1) Because there is no need for integrals, it is comparatively easy to program. 2) The points can easily be chosen to improve the accuracy (we use a distribution with more points in low-energy regions); a uniform or regular grid is not necessary. 3) It can be used with any basis functions. 4) Unlike some collocation or pseudospectral methods, it is straightforward to use more points than basis functions.

To put the accuracy of our computed energies in perspective, consider energies obtained with the LSE method, a type of collocation approach. The LSE ground state energies are excellent,[19,56] but the error is about $10^{-2}$ *a.u.* for the $n = 2$ level of the H atom.[19] An accuracy of $10^{-2}$ *a.u.* was obtained with the LSE (compared to full CI) for the four lowest states of LiH using $10^7$ points.[56] These larger errors may be due to the fact that the LSE basis favors the ground state. We demonstrate that with very simple basis functions good accuracy can be achieved for many levels.

In a recently published article,[57] Jerke and Poirier (JP) solve the ESE using low-rank basis functions represented in CP format[58] generated by evaluating matrix-vector products and using a sum-of-products representation of the Hamiltonian. Their method is similar to the reduced rank block power method of Leclerc and Carrington.[59] JP show that their low-rank method works well for the ESE for the H and He atoms and $H_2$. The accuracy of our and their energies is comparable. We have, so far, applied rectangular collocation only to 3D electronic problems, but we do test it for Kohn-Sham equations. The Kohn-Sham equations are hugely important for studying properties of molecules and materials.

Much can be done to improve and extend the method we present in this paper. The accuracy of the energies presented here should be regarded as a sort of lower bound on performance. It ought to be possible to optimize further both the basis functions and the point distribution. For 3D problems, the multi-$\zeta$ basis we used is efficient and natural. It exploits spherical symmetry. To go beyond 3D, one can probably use basis functions related to those used with variational methods. Decreasing the basis size permits using fewer collocation points.[6,11] Although computing Coulombic potential values is cheap, computing values of the DFT effective potential is more costly. In previous papers in which we use rectangular collocation for the vibrational Schrödinger equation, we showed that optimizing basis parameters and collocation point placement can significantly improve the accuracy.[4,6-9,11] As the basis quality improves, the choice of collocation points becomes less critical. Optimization was not done in this paper, nor did we ensure the regularity of the grid or symmetry which would be needed to reproduce exactly the degenerate states, both could be implemented.

## 5 Acknowledgements

Aditya Kamath is thanked for helping with parts of the code.

## 6 Conflicts of interest

There are no conflicts of interest to declare.